\documentclass[11pt]{article}
 \newtheorem{theorem}{Theorem}[section]   
       \newtheorem{example}{Example}[section]       
   \newtheorem{lemma}{Lemma}[section]      
    \newtheorem{corollary}{Corollary}[section]      
    \newtheorem{definition}{Definition}[section]

      \newtheorem{remark}{Remark}[section]      
      \newcommand{\finishproof}{ \hspace*{\fill}{$\Box$}\medskip}
\usepackage{latexsym}
\usepackage{amsmath}
\usepackage{gastex}
\usepackage{color}
 \usepackage{amsfonts,amssymb,amscd,latexsym,graphicx,color}

 \title{Uniform Interpolation for Propositional and Modal Team  Logics }
\author{Giovanna D'Agostino  \\
Department of Math., Comp. Science, and Physics,\\
University of Udine,
 Via delle Scienze 206, 
 33100 Udine, Italy\\
 {\tt giovanna.dagostino@uniud.it}\\
}              
 \date{}

\begin{document}

 \maketitle

\abstract{In this paper we discuss the property  of  uniform interpolation  in Propositional and  Modal Team  Logics}

\section{Introduction}
Interpolation   is a desirable  property for a logic. In very general terms it states that if a formula $G$   is a consequence  of  a formula $F$, then only the common language between the two  formulas is important,  because  $G$ is   also  a  consequence of a formula in the common language.  
Uniform interpolation is stronger than Craig interpolation, in the sense that the  interpolant between $F$ and $G$ only depends on $F$ and on the common language between $F$ and $G$, but not on $G$ itself:  in a logic enjoying uniform interpolation, given a sublanguage   $L$ of the language of $F$, if $G_1,G_2$ are two consequences of $F$   having $L$ as common language with $F$, then both $G_1,G_2$ will be logical conseuqnce of the uniform interpolant of $F$ w.r.t. $L$. 

Although uniform interpolation is a stronger property than Craig interpolation,  it is, in some way,  more stable. Suppose we have two logics, $L_1,L_2$ where  $L_1$  is  more expressive than $L_2$, and enjoys  Craig interpolation.  Hence, if $\phi, \psi$ are $L_2$ formulas with $\phi\models \psi$, then $\phi, \psi$  have an interpolant in $L_1$, which, however, could be a formula which is not equivalent to a formula in  $L_2$.  On the other hand,  we will prove that  $\mathcal {FPTL}$ enjoys uniform interpolation and we  that the  uniform interpolant  of a $\mathcal {PDEP}$ formula is equivalent to a $\mathcal {PDEP}$ formula.  More generally, if  $L_1,L_2$ are as above,  the  logic $L_1$    enjoys  uniform interpolation,  and  $L_2$ has some nice semantical characterization inside $L_1$, we can  try to prove that the $L_1$ uniform interpolant of a formula  in $L_2$ is  (equivalent to) an  $L_2$-formulas. 

As we shall see, this strategy proved to be quite fruitful for team logics.

\section{Propositional and Modal  Team  Logics}

In the sequel   $Prop$   denotes a nonempty set of propositional letters.
A team $X$ over $Prop $ is a set of valuations, where a valuation $s$ is  a function 
$s: Prop\rightarrow \{0,1\}.$

Team formulas are built from literals  $p, \neg p$ (for $p\in Prop$) and the constant $\bot$ using  conjunction $\wedge$ and the team disjunction $\otimes$,  and are interpreted on teams as follows:

 \begin{align*}
 X&\models  \bot   &\Leftrightarrow  \qquad &  X=\emptyset\\
 X &\models  p_i  &\Leftrightarrow  \qquad & s(p_i)=1\   \hbox{\ for all \ } s\in X\\
   X &\models   \neg p_i  &\Leftrightarrow  \qquad & s(p_i)=0\   \hbox{\ for all \ } s\in X\\ 
      X&\models    \phi_1\wedge \phi_2 &\Leftrightarrow  \qquad &  X\models \phi_1\hbox{\ and  \ }   X\models\phi_2 \\
   X&\models \phi_1\otimes\phi_1 &\Leftrightarrow  \qquad & \exists X_1, X_2 ~~ X=X_1\cup X_2, ~ X_1\models \phi_1, ~ X_2\models \phi_2. \\
 \end{align*}
 
We  use  the constant $\top$   to denote the formula $p\otimes\neg p$,  for a propositional variable $p$ in the language; notice that $\top$ is true in any team. 
Moreover, we    consider also the non empty disjunction $\circledast$,  classical  disjunction $\vee$, and  a constant for non-emptiness $NE$:
  \begin{align*}
   X&\models \phi_1\circledast \phi_2 &\Leftrightarrow  \qquad & X=\emptyset \hbox{~or~} \exists X_1\neq \emptyset, X_2\neq \emptyset, ~X=X_1\cup X_2, \\
   &&& X_1\models \phi_1, ~ X_2\models \phi_2\\
  X&\models \phi \lor \psi  &\Leftrightarrow  \qquad &  X\models \phi \hbox{~or~} X\models \psi \\
  X &\models NE  &\Leftrightarrow  \qquad & X\neq \emptyset.
      \end{align*}

Next, we add the modal operators:

\begin{definition}  The formulas of Modal  Logic   ${\mathcal M}{\mathcal L}$ are defined by
\[
\alpha:=  ~~   p~|~ \neg p ~|~ \bot  ~| ~ \alpha_1\wedge \alpha_2~|~ \alpha_1\otimes \alpha_2      ~|~  \Diamond \alpha_1~|~\Diamond  \alpha_2 \]
  where $p\in Prop$.
\end{definition}

To interpret the  modality operators $\Box\phi, \Diamond \phi$   we  enrich the semantics by considering teams $X$ as subsets of the set of words of a Kripke model, defined, as usual, 
as  a tuple $M=(W, R, V)$, where  $W$ is a non empty  set,  $R\subseteq W\times W$ is the {\em accessibility} relation,  and  $V:W\rightarrow Pow(Prop)$.

We use the following notation for  $X,Y\subseteq W$:
\[R(X):= \{s\in W~:~ \exists t \in X ~sRt\};\] 
\[XR Y \Leftrightarrow \forall x\in X \exists  y \in Y~ x Ry \wedge \forall y \in Y \exists x \in X~ xRy.\]

\begin{definition}  (Modal   Team Semantics) ~ If $M= (W,R,V)$  is a Kripke model and $X\subseteq W$ is a team in $M$ then the semantics of   ${\mathcal M}{\mathcal L}$ is defined as follows
 \begin{align*}
  (M,X) &\models p  &\Leftrightarrow  \qquad &  s\in V(p),   \hbox{\ for all \ } s\in X\\
   (M,X)&\models \neg p  &\Leftrightarrow  \qquad &  s\not \in V(p),   \hbox{\ for all \ } s\in X\\ 
     (M,X) &\models \bot &\Leftrightarrow  \qquad & X=\emptyset\\
     (M,X)&\models   \alpha_1\wedge \alpha_2 &\Leftrightarrow  \qquad & (M,X)\models \alpha_1\hbox{\ and  \ }  (M,X)\models\alpha_2 \\
   (M,X)&\models\alpha_1\otimes\alpha_1 &\Leftrightarrow  \qquad & \exists X_1, X_2 ~~ (M, X_1)\models \alpha_1,~(M,X_2)\models \alpha_2\\
   &&&  \hbox{~and ~} X=X_1\cup X_2 \\ 
  (M,X)&\models   \Diamond  \alpha\  &\Leftrightarrow  \qquad & \exists Y ~  XRY \hbox{\ and  \ }  (M,Y)\models \alpha\\
  (M,X)&\models   \Box\alpha\  &\Leftrightarrow  \qquad & (M, R(X)) \models \alpha\\
 \end{align*}
\end{definition}

If the team $X$  is a singleton, then  the semantics of ${\mathcal M}{\mathcal L}$ formulas coincides   with the classical modal semantics on Kripke model, with $\otimes$ behaving as a standard disjunction. If $\phi \in {\mathcal ML}$ we call {\em singleton semantics} the usual semantics of modal logic, where formulas are interpreted over  pointed Kripke models $(M,w)$ and   not over teams $(M,X)$.

\subsection{Dependence, Inclusion, Independence}\label{modal_fragments}

Team semantics allow us to consider various    notions of dependence between data. 
To this end,  new \lq atoms\rq~  are added to the basic framework discussed above (both in the propositional and in the modal case). In this paper we consider  dependence atoms $_=(\overline \alpha,    \gamma)$, inclusions atoms  $\overline \alpha \subseteq  \overline \alpha'$, independence atoms $\overline \alpha\bot \overline \beta$,  where $\overline \alpha, \overline \alpha', \overline \beta $    are sequences of formulas in $\mathcal {M L}$ (with  $\alpha, \alpha'$ of  the same lenght), and $\gamma\in \mathcal {M L}$.  

\begin{definition} \cite{Kontinen2015,Kontinen2017,Hella2015,Hella2014}\\ \
 The formulas of Modal Dependence    Logic    $\mathcal {MDL}$ are defined by
\[
\phi:=  ~~   p~|~ \neg p ~|~ \bot  ~| ~   \phi_1\wedge \phi_2~|~ \phi_1\otimes \phi_2  ~|~_=(\overline \alpha,  \gamma)     ~|~  \Diamond \phi ~|~\Box \phi \]
  where $p\in Prop$, $\overline \alpha=\alpha_1, \ldots, \alpha_h$,   and $\alpha_i, \gamma$ are formulas in   $\mathcal {ML}$.\\
  
 The formulas of Modal Inclusion    Logic    $\mathcal {MINC}$ are defined by
\[
\phi:=  ~~   p~|~ \neg p ~|~ \bot  ~| ~   \phi_1\wedge \phi_2~|~ \phi_1\otimes \phi_2  ~|~\overline \alpha\subseteq  \overline \alpha'     ~|~  \Diamond \phi~|~\Box  \phi \]
  where $p\in Prop$, $\overline \alpha=\alpha_1, \ldots, \alpha_h$, $\overline \alpha'=\alpha_1', \ldots, \alpha_h$ and $\alpha_i, \alpha_j'\in \mathcal {M  L}$.   \\
  
  The formulas of Modal Independence   Logic    $\mathcal {MIND}$ are defined by
\[
\phi:=  ~~   p~|~ \neg p ~|~ \bot  ~| ~   \phi_1\wedge \phi_2~|~ \phi_1\otimes \phi_2  ~|~\overline \alpha\bot \overline \beta     ~|~  \Diamond \phi~|~\Box  \phi \]
  where $p\in Prop$, $\overline \alpha=\alpha_1, \ldots, \alpha_h$, $\overline \beta =\beta_1, \ldots, \beta_k$ and $\alpha_i, \beta_j\in \mathcal {M  L}$.   \\
  \end{definition}
  
  Notice that the new atoms  $_=(\overline \alpha,    \gamma)$,  $\overline \alpha \subseteq  \overline \alpha'$, $\overline \alpha\bot \overline \beta$ are defined only on  $\mathcal {ML}$ formula  and cannot be nested.  By convention, we use letters $\alpha, \beta, \ldots$ to denote $\mathcal {ML}$ formulas, and letters $\phi, \psi, \ldots$  to denote    formulas  in $\mathcal {MDL},  \mathcal {MINC}, \mathcal {MIND}$.

  To give a semantics to  the new atoms, 
given   a Kripke   model    $M=(W,R,V)$ and $s\in W$,  we  consider    the valuation  functions $M_s$ on   $ \mathcal {M  L}$-formulas    defined  as follows:
  \[M_s(\alpha)  =\begin{cases} 1, ~\hbox{if}~ M,s\models \alpha;\\
  0 , ~\hbox{if}~ M,s\not \models \alpha,
  \end{cases}\]
  \noindent where $M,s\models \alpha$ denote the usual (singeton) semantics for modal formulas. 
Moreover, if $\overline \alpha=\alpha_1, \ldots, \alpha_h$  is a sequence of formulas in $\mathcal {M  L}$ we define $M_s(\overline \alpha) = (M_s(\alpha_1), \ldots, M_s(\alpha_h))$.

\begin{definition} The semantics of the new   atoms is defined over a Kripke team model $(M,X)$ as follows:
  \begin{align*}
(M,X) \models  ~_=(\overline \alpha,   \beta)  &~\Leftrightarrow~ &&\\
 & \forall s,s' \in X~(M_s (\overline \alpha)=   M_{s'} (\overline \alpha)   \Rightarrow  M_s(\beta)=  M_{s'}(\beta));\\
 (M, X)  \models  ~\overline \alpha \subseteq  \overline \alpha' &~\Leftrightarrow~ && \\
 & \forall s \in X ~\exists s' \in X ~~ M_s(\overline \alpha)= M_{s'}(\overline \alpha');\\
 (M, X)  \models ~  \overline \alpha \bot  \overline \beta  &~\Leftrightarrow ~&& \\
 & \forall s \forall s'  \in X ~\exists s''(M_{s''}(\overline \alpha)=M_s(\overline \alpha) \wedge M_{s''}(\overline \beta)=M_{s'}(\overline \beta)).
  \end{align*}
  \end{definition}

  In correspondence  to any Modal Team Logic we also consider its propositional  variant, which  has the same syntax except for the absence of the modal operators. Hence, we consider Propositional Dependence Logic, Propositional Inclusion Logic, and Propositional Independence Logic.  In the propositional case, the new  atoms   are evaluated over a set $X$ of valuations $s: Prop \rightarrow \{0,1\}$,  as expected: e.g.    
  \[X \models  ~_=(\overline \alpha,   \beta)   \Leftrightarrow  \qquad  \forall s,s' \in X~(s (\overline \alpha)=  s' (\overline \alpha)   \Rightarrow  s(\beta)=  s'(\beta)).\]

 In the following table we   list  all     team logics we will   consider in this paper:
 
 \newpage 
 \begin{table}
\begin{tabular}{|c|c|c|c|}  
\hline
LOGIC&ATOMS&CONN.&MOD. OPER.\\
\hline
Classical Prop.Team Logic $\mathcal{CPL}$ &$p_i, \neg p_i, \bot$ & $\wedge, \otimes$  & $-$\\
\hline
 Prop. Dependence Logic $\mathcal{PDEP}$ &$p_i, \neg p_i, \bot, _=(\overline \alpha,   \beta)$ & $\wedge, \otimes$  & $-$\\
 \hline
 Prop. Inclusion  Logic $\mathcal{PINC}$ &$p_i, \neg p_i, \bot,  \overline \alpha \subseteq  \overline \beta$ & $\wedge, \otimes$  & $-$\\
 \hline
 Prop. Independence Logic $\mathcal{PIND}$ &$p_i, \neg p_i, \bot, \overline \alpha \bot \overline \beta$ & $\wedge, \otimes$  & $-$\\
\hline
 Full Prop.Team Logic  $\mathcal{FPTL}$ &$p_i, \neg p_i, \bot, NE$ & $\wedge, \otimes, \lor $  & $-$\\
 \hline 
Modal Team Logic $\mathcal{ML}$&$p_i, \neg p_i, \bot$ & $\wedge, \otimes$  & $\Box$, $\Diamond$\\
\hline
Modal Dependence Logic $\mathcal{MDEP}$ &$p_i, \neg p_i, \bot, _=(\overline \alpha,   \beta)$ & $\wedge, \otimes$  & $\Box$, $\Diamond$\\
 \hline
Modal Inclusion  Logic $\mathcal{MINC}$ &$p_i, \neg p_i, \bot,  \overline \alpha \subseteq  \overline \beta$ & $\wedge, \otimes$  & $\Box$, $\Diamond$\\
 \hline
Modal Independence Logic $\mathcal{MIND}$ &$p_i, \neg p_i, \bot, \overline \alpha \bot \overline \beta$ & $\wedge, \otimes$  & $\Box$, $\Diamond$\\
\hline
 Full Modal Team Logic  $\mathcal{FMTL}$ &$p_i, \neg p_i, \bot, NE$ & $\wedge, \otimes, \lor $  & $\Box$, $\Diamond$\\
 \hline
\end{tabular}
 \caption{A list of (Modal) Team Logics \label{list}}
\end{table}

When $L$ is one of the logic above,  we   denote by $(L, \models)$ the pair consisting  of  the logic and its consequence relation, defined as usual.

%
%
%
  
  \section{Uniform Interpolation in the Propositional Team Context}\label{sec2}

 If $\phi$ is a formula of a  team logic we  denote by  ${\cal L}(\phi) \subseteq Prop$ the finite set of  proposition  from which  $\phi$ is constructed. 

\begin{def}\label{int1}
Let $\phi$ and $\psi$ be two formulas in  a team  logic $(L, \models)$ such 
that $\phi \models \psi$. 
Then $\theta$ is an 
{\em interpolant} of $\phi,\psi$ 
iff:
\begin{enumerate}
\item $\phi \models \theta$ and $ \theta \models \psi$;
\item ${\cal L}(\theta ) \subseteq {\cal L}(\phi ) \cap 
                                   {\cal L}(\psi )$.
   \hspace*{\fill}$\Box$
\end{enumerate}
\end{def}

\noindent In words: if $\phi \models\psi$,  an interpolant of  $\phi,\psi$ is a 
formula in the common language of $\phi$ and $\psi$ 
which sits in between $\phi$ and $\psi$. 

\begin{definition}
Given a formula $\phi$ and a language 
${\cal L}^\prime \subseteq {\cal L}(\phi )$, 
the {\em uniform interpolant} of 
$\phi$ with respect to ${\cal L}^\prime$ 
is a formula $\theta$ such that:
\begin{enumerate}
\item $\phi \models \theta$;
\item Whenever $\phi \models  \psi$ and
      ${\cal L}(\phi ) \cap {\cal L} (\psi ) \subseteq {\cal L}^\prime$
      then $\theta\models \psi$.
\item ${\cal L}(\theta ) \subseteq {\cal L}^\prime$. 
 \hspace*{\fill}$\Box$
\end{enumerate}
\end{definition}

\noindent 

When we say that  a  logic has (uniform)
interpolation we mean that we can always find a (uniform)
interpolant when the appropriate conditions are satisfied. 
Clearly, if  a logic has uniform interpolation,
it also enjoys Craig interpolation. For if $\phi\models \psi$, 
simply choose ${\cal L}^\prime = {\cal L}(\phi )\cap {\cal L}(\psi )$.
The interpolant  between $\phi$ and $\psi$ is then the uniform interpolant of $\phi$
relative to ${\cal L}^\prime$. This explains why we call this
formula a {\em uniform} interpolant: no information is needed
about the formula $\psi$ except which non-logical symbols it has
in common with $\phi$.   

\medskip

Before proving uniform interpolation in the modal  team context,   we recall  the easy proof of uniform interpolation for Classical Propositional Logic, and show that 
it cannot be applied to the propositional  team context, except for  the case of $\mathcal{CPL}$ (which is the simpler  logic with team semantics we shall consider). 

In   Classical Propositional (singleton) Logic\footnote{that is: propositional logic with the usual semantics, which coincides with the team semantics ristrected to singleton teams}  it is well know (and easy to prove) that the formula  $\phi[p|\top]\vee \phi[p|\bot]$ is a    uniform interpolant  for $\phi$ with respect to  ${\cal L}(\phi)\setminus \{p\}$. Moreover,  we can iterate this construction to obtain a uniform interpolant with respect to  any subset of   ${\cal L}(\phi)$. This immediately  implies  that Classical Propositional Team  Logic $\mathcal{CPL}$ enjoys uniform interpolation,  the uniform interpolant of a propositional formula 
$\phi(p)$ with respect to ${\cal L}(\phi)\setminus \{p\}$ being again   $\phi[p|\top]\otimes \phi[p|\bot]$.

 Note that all formulas in Classical Propositional Team  Logic are  downward closed, union closed,  and local,  where a formula $\phi$  is:
 
 \begin{definition}\label{properties} (see \cite{YangV17})
\begin{enumerate}
\item local, if $X\models \phi$ and $Y=_{L(\phi)} X$ implies $Y\models \phi$, 
 \begin{eqnarray*} (\hbox{where ~}  Y=_{{\cal L}(\phi)} X~ \Leftrightarrow ~&\forall s\in Y\exists s' \in X s(p)=s'(p), \forall p \in  {\cal L}(\phi)\\
&\forall s\in X\exists s' \in Y s(p)=s'(p), \forall p \in  {\cal L}(\phi));
\end{eqnarray*}
\item downward closed,   if  $X\models \phi$ and $Y\subseteq X$ implies $Y\models \phi$;
 \item   union closed, if $X_1\models \phi$ and $X_2\models \phi$ implies $X_1\cup X_2\models \phi$.
 \end{enumerate}
 \end{definition}
 We next show how these properties play a separate role in order to ensure  that     the formula   $\phi[p|\top]\otimes \phi[p|\bot]$  is a  a uniform interpolant for a formul  $\phi$  in $\mathcal{CPL}$,  with respect to ${\cal L}(\phi)\setminus \{p\}$. 
  To this end, we have first to recall some lemma on substitutions in a team context. 
  
  \medskip
  
Given a team $X$, we  define  \[X[p|\top]:=\{ s[p|\top]:s\in X\}, ~ X[p|\bot]:=\{ s[p|\bot ]:s\in X\},\] where 
\[ s[p|\top](q)=
 \begin{cases}
 s(q) \hbox{~ if ~} q\neq p\\
1 \hbox{~ if ~} q= p
 \end{cases} ~~
  s[p|\bot](q)=
 \begin{cases}
 s(q) \hbox{~ if ~} q\neq p\\
0 \hbox{~ if ~} q= p
 \end{cases}
 \]
 
In general, team logics do not have a  good notion of substitution, unless we restrict to  classical substitutions.  In particular, if  we  define $\phi[p|\top], \phi[p|\bot]$ by induction,  as usual:
\[\phi[p|\top]:=
\begin{cases}
\top, \hbox{~ if~} \phi=p;\\
\bot, \hbox{~ if~} \phi=\neg p ~ \hbox{or}~ \phi=\bot;\\
\phi_1[p|\top]\circ \phi_1[p|\top],\hbox{~ if~} \phi=\phi_1\circ \phi_2, \circ\in\{\otimes, \vee, \wedge\};\\
\overline \alpha[p|\top]\subseteq \beta[p|\top] \hbox{~ if~} \phi=\overline \alpha\subseteq\beta;\\
\vdots
\end{cases}\]
\[\phi[p|\bot]:=
\begin{cases}
\bot, \hbox{~ if~} \phi=p;\\
\top, \hbox{~ if~} \phi=\neg p;\\
\phi_1[p|\bot]\circ \phi_1[p|\bot],  \hbox{~ if~} \phi=\phi_1\circ \phi_2, \circ \in\{\otimes, \vee, \wedge\};\\
\overline \alpha[p|\bot]\subseteq \beta[p|bot] \hbox{~ if~} \phi=\overline \alpha\subseteq\beta;\\
\vdots
\end{cases}~~
\]
 then  the syntactic substitution reflects on the semantics side as follows  (see  \cite{YangV17}):   
\begin{lemma}\label{substitutions} If $\phi$ is a propositional team formula and $X$ is a team then
\[X[p|\top]\models \phi \Leftrightarrow X\models \phi[p|\top];~~X[p|\bot]\models \phi \Leftrightarrow X\models \phi[p|\bot].\] 
In particular, if $X\models p$, then 
$X\models \phi \Leftrightarrow X\models \phi[p|\top].$\\
Similarly,  if  $X\models \neg p$, then 
$X\models \phi \Leftrightarrow X\models \phi[p|\bot].$
\end{lemma}
 
\begin{lemma}  If  $\phi$ is downward closed then 
$\phi \models  \phi[p|\top] \otimes  \phi[p|\bot]$. 
\end{lemma}
{\bf Proof}  Suppose $X\models  \phi $.    Consider $X_0=\{s\in X: s(p)=0\}$ and  $X_1=\{s\in X: s(p)=1\}$. By downward closure,   $X_0\models  \phi$  and hence $X_0\models  \phi[p|\bot]$  by the observation above. Similarly,  $X_1\models  \phi[p|\top]$, and $X \models \phi[p|\top] \otimes  \phi[p|\bot]$ follows.
\finishproof
 
However, if $\phi$ is not downward closed, the previous lemma does not hold, as the following example shows. 
 
 \begin{example}
\label{ex1}  If 
 $\phi:=~ (p\wedge q) \circledast (\neg p \wedge q) $, then $\phi[p|\top] \otimes  \phi[p|\bot]$ is not a logical consequence of  $\phi$. In particular, $\phi[p|\top] \otimes  \phi[p|\bot]$ is not a uniform interpolant of $\phi$. 
 \end{example}
{\bf Proof} ~
$\phi[p|\top]= ( \top \wedge q ) \circledast (\bot  \wedge q)$, and  $\phi[p|\bot]= ( \bot  \wedge q ) \circledast (\top \wedge q)$;  both formulas are  easily seen to be  true only  for the empty team, 
hence  $X\models \phi[p|\top] \otimes  \phi[p|\bot]$ implies $X=\emptyset$. 
On the other hand,  $\phi$  is satisfied by the non empty team $X=\{s_1,s_2\}$ with 
  $s_1(p)=s_1(q)=s_2(q)=1$, and $s_2(p)=0$.
  Hence,  $\phi[p|\top] \otimes  \phi[p|\bot]$ is not a logical consequence of  $\phi$. 
  \finishproof

\begin{lemma}  \label{easyexistsp} Suppose $\phi $ is  union closed  and $\psi$ is a  local  formula   such that $\phi\models \psi$ and $p\not \in {\cal L}(\psi)$. Then 
\[  \phi[p|\top] \otimes  \phi[p|\bot] \models  \psi\]
\end{lemma}
{\bf Proof}~
Suppose $X\models \phi[p|\top] \otimes  \phi[p|\bot]$. Then there are $X_1,X_2$ such that $X=X_1\cup X_2$ and 
$X_1\models  \phi[p|\top]$, $X_2\models  \phi[p|\bot]$. 
By Lemma  \ref{substitutions} we obtain $ X_1[p|\top]\models  \phi$, and $ X_2[p|\bot]\models  \phi$, and, by   union closure,   $X_1[p|\top]\cup X_2[p|\bot]\models \phi$. 
From $\phi \models \psi$ we obtain that $X_1[p|\top]\cup X_2[p|\bot]\models   \psi$,  and hence $X\models \psi$, since $\psi$ is local and  \[X=_{ L(\psi)\setminus \{p\}}(X_1[p|\top]\cup X_2[p|\bot]).\]  
\finishproof

 If $\phi$ is not union  closed, the previous lemma does not hold, as the following example shows. 
 \begin{example}\label{ex2} 
 Consider the formula 
 \[ \phi(p,q):= ~  _=(p,q)\wedge  {_=}(p)\]
 
 We have:
 \[ \phi[p|\top] \equiv ~_=(\top, q) \wedge   {_=}(\top) \equiv ~_=(q)\equiv ~ \phi[p|\bot] \]
 Hence, $\phi[p|\top] \otimes \phi[p|\bot]\equiv {_=}(q) \otimes {_=}(q) \equiv \top$.
 On the other hand, it is clear that $\phi \models {_=}(q)$, although $\top\not \models {_=}(q)$.
It follows that  the formula $\phi[p|\top] \otimes \phi[p|\bot]$ is not a uniform interpolant for $\phi$ with respect to ${\cal L}(\phi)\setminus\{p\}$.
 
\end{example}
Hence, if  we consider a propositional team logic which is not downward closed or not union closed,  we cannot prove uniform interpolation using the formula $\phi[p|\top] \otimes  \phi[p|\bot]$.  On the other hand, one can easily check that  in Example (\ref{ex2}) the formula ${_=}(q)$ is the correct uniform interpolant in any local logic containing $\phi$:  one can easily verify that $\phi\models  {_=}(q)$; moreover,   suppose  $\phi \models \psi$ with $p\not \in {\cal L}(\psi)$, and $X\models {_=}(q)$;  then if  $Y:=X[p|\top]$ we have $Y\models \phi$ and hence $Y\models \psi$; but then $X\models \psi$ because  $Y=_{{\cal L}(\phi)} X$ and $\psi$ is local. 

As we shall see, at least for   Propositional Dependence Logic and for Propositional Inclusion  Logic we can prove uniform interpolation. 
Similarly,   uniform interpolation for Modal Team Logic can be easily proved from uniform interpolation of  standard  modal logic, but this easy proof  cannot be used for other team modal logics, where we have to use other means. 

 \section{Expressiveness of   Team Logics}

Given a formula $\phi$ in a propositional logic with team semantics we denote by $||\phi||$ the class of team models of $\phi$:
\[
||\phi||=\{X: X\models \phi\}.
\]
If $L$ is a fragment of Full Propositional Team Logic $\mathcal{FPTL}$, we denote by $Team_L$ the class of team models of formulas in $L$:
\[Team_L=\{||\phi||:  \phi\in L\}\]
 
 We say that a team  propositional logic $L$ is expressively complete for a class of team properties $\mathcal X$ if, 
 \[\mathcal X=Team_L, \] 
 that is:
for every formula $\phi\in L$,  the set of team satisfying $\phi$ belongs to $\mathcal X$ and, moreover, every team property belonging to  $\mathcal X$  coincides with the set of team satisfying a formula in $L$.  We have: 
 
\begin{lemma}\cite{YangV17} 
\begin{enumerate} 
\item $\mathcal{CPL}$ is  expressively complete for the class of   non empty, downward and union closed team properties;
\item $\mathcal{PDEP}$ is  expressively complete for the class of   non empty,  downward closed team properties;
\item $\mathcal{PINC}$ is  expressively complete for the class of   non empty,  union closed team properties;
\item $\mathcal{FPT}$ is  expressively complete for the class of  all  team properties.
\end{enumerate}
 \end{lemma}

In order to state an analogous lemma for modal team logics, we consider modal team properties as  sets of modal team models    $(M,X)$, and define:
  
\begin{definition} A   team property   ${\mathcal C}$ is:
\begin{enumerate}
\item {\em downward closed: }   $(M,X)\in {\mathcal C}$  and $Y\subseteq X$   implies $(M, Y) \in {{\mathcal C}}$;
\item {\em union closed:}  $(M,X_i)\in {{\mathcal C}}$   implies  $(M,\bigcup_iX_i )\in   {{\mathcal C}}$.
\end{enumerate}
\end{definition}

Moreover, to state the expressiveness results,  we need the notion of team (bounded) bisimulation, a generalization of the usual notion of  (bounded) bisimulation: 
 \begin{definition}
 If   $M=(W,R, V), M'=(W',R', V')$ are Kripke models, $(w,w')\in M\times M'$, and $k\in \mathbb N$,     we say that $(M,w), (M',w')$ are  $k$-bisimilar  (notation: $(M,w) \rightleftharpoons^k (M',w')$), iff  for all $i\leq k$ there is  $B_i\subseteq M\times M'$ such that $(w,w')\in B_k$, and  for all $(v,v')\in B_{i+1}$ it holds:
\begin{enumerate}
\item  $V(v)=V(v')$;
\item if $vRu$ there exists $u'$ such that $v'R'u'$ and $(u,u')\in B_i$;
\item if $v'Ru'$ there exists $u$ such that $vRu$ and $(u,u')\in B_i$.
\end{enumerate}
We say that $(M,w), (M',w')$ are   bisimilar  (notation: $(M,w) \rightleftharpoons  (M',w')$), iff  there exists  $B\subseteq M\times M'$ such that $(w,w')\in B$, and  for all $(v,v')\in B$ it holds:
\begin{enumerate}
\item  $V(v)=V(v')$;
\item if $vRu$ there exists $u'$ such that $v'R'u'$ and $(u,u')\in B$;
\item if $v'Ru'$ there exists $u$ such that $vRu$ and $(u,u')\in B$.
\end{enumerate}

 \end{definition}
 \medskip

We shall also consider bisimulations and bounded bisimulation where   condition $1.$ above is  resticted to a subset  $\cal P$ of propositions, that is, we require that  $V(v)\cap {\cal P}=V(v')\cap {\cal P}$.
Two models $(M,w), (N,v)$ which are bisimilar  ($k$-bisimilar) w.r.t. the propositions in $\cal P$  are denoted by $(M,w)\rightleftharpoons_{\cal P}(M',w')$, 
($(M,w)\rightleftharpoons^k_{\cal P}(M',w')$,  respectively). 

 \medskip

The notion of bisimulation is extended to team models as follows.   
\begin{definition} 
Let  $\cal P$  be a set of propositional variables.  The team models $(M,X), (N,Y)$ are  $\cal P$-bisimilar   if
\[\forall x\in X~\exists y \in Y (M,x)\rightleftharpoons_{\cal P} (N,y)  \hbox{~and~}\forall y\in Y~\exists x \in X (M,x)\rightleftharpoons_{\cal P} (N,y).\]
Team bisimilar models are denoted by   $(M,X)\rightleftharpoons_{\cal P} (N,Y)$. 
\end{definition}
\begin{remark}\label{max}
By considering the maximal bisimulation between models, if  $(M,X)\rightleftharpoons_{\cal P} (N,Y)$ we may suppose w.l.o.g. that there exists a ${\cal P}$-bisimulation $B$ between $M$ and $N$ such that 
\[\forall x\in X~\exists y \in Y (x,y)\in B \hbox{~and~}\forall y\in Y~\exists x \in X (x,y)\in B.\]
\end{remark}

One can easily prove that all formulas $\phi$  in the logics listed on table \ref{list}. are invariant under bisimulation, that is, 
\[(M,X)\models \phi ~\hbox{ and }~  (M,X)\rightleftharpoons_{\cal L(\phi)} (N,Y) ~\hbox{ implies}~ (N,Y)\models \phi.\] 
Similarly, if $md(\phi)$, the modal depth of $\phi$, is defined as the maximal number of nested modal operators in $\phi$,  we have:
\[(M,X)\models \phi ~\hbox{ and }~  (M,X)\rightleftharpoons^k_{\cal L(\phi)} (N,Y) ~\hbox{ implies}~ (N,Y)\models \phi.\] 
 
 The expressiveness results for modal team logics are based on the following definition:
 
 \newpage
 \begin{definition} \ \\
 \begin{itemize}
 \item A class $\cal K$ of team models is    {\em bisimulation invariant}  if 
 \[ (M,X)\in {\cal K } \hbox{~ and~} (M,X)\rightleftharpoons (N,Y)  \hbox{~ implies~}  (N,Y) \in {\cal K }.\]
\item  A class $\cal K$ of team models is  {\em  first order definable} if there exists a first order  formula $\phi(V)$ with a monadic  variable $V$ in the language 
$\{R, =, P_1, \ldots, P_n,\ldots\}$, 
where $R$ is a binary relational symbol representing the accessibilty relation and $P_1, \ldots, P_n,\ldots$ are unary relational symbol representing the propositions, such that, for all team model $(M,X)$ it holds:
\[ (M,X) \in {\cal K} \Leftrightarrow   M, V:=X \models \phi(V),\]
where   $M, V:=X $ on the right     $M$ is considered  as a first order model for the language $\{R, =, P_1, \ldots, P_n, \ldots \}$,   and we interpret the monadic  variable $V$ by the set $X$. 
\end{itemize}
\end{definition}

Given a formula $\phi$ in a modal logic with team semantics we denote by $||\phi||$ the class of team models of $\phi$:
\[
||\phi||=\{(M,X): (M,X)\models \phi\}.
\]
As in the propositional case, we say that a team modal logic $L$ is expressively complete for a class of team properties $\mathcal X$ if 
 \[\mathcal X= Team_L,\]
where $Team_L= \{||\phi||: \phi \in L\}$.  
 We have:

\begin{theorem}\label{Kontinen} \cite{Hella2014, Hella2015, Kontinen2015}
\begin{enumerate}
\item $\mathcal{ML}$ is  expressively complete for the class of all first order definable,   non empty,   downward and union closed,   bisimulation invariant team properties;
\item $\mathcal{MDEP}$ is  expressively complete for the class of all first order definable,  non empty,  downward closed,   bisimulation invariant team properties;
\item $\mathcal{MINC}$ is  expressively complete for the class of  all first order definable,  non empty,  union closed,  bisimulation invariant team properties;
\item Full Modal Team Logic $\mathcal {FMTL}$ is expressively complete for  the class of all   first order definable,  bisimulation invariant team properties.
\end{enumerate}
 
\end{theorem}

\section{Amalgamation}

To prove  uniform interpolation  we need  the notion of amalgamation, defined below. 

\begin{lemma} \label{kripke_amalgamation}  Let ${\cal P}, {\cal Q}$ are sets of propositions and  let $B$ be a ${{\cal P}\cap {\cal Q}}$-bisimulation between $M,N$ with $B\neq \emptyset$. 
The  $B$-{\em amalgamation}  $K$ of   $M,N$, is a Kripke model over the propositions  ${\cal P}\cup {\cal Q}$   defined as follows:\\
-The domain of $K$ is the relation  $B$. \\
- The accessibility relation $R^K$ is given by 
\[(m,n) R^K (m',n') ~\Leftrightarrow~ mR^Mm' \hbox{~ and~} n R^N n'.\]
For all propositional variables $r\in {\cal P}\cup {\cal Q}$ and $(m,n)\in B$ we have 
\[(m,n)\models r ~\Leftrightarrow ~ 
\begin{cases}
(M,m)\models r, \hbox{\ if \ } r\in \cal P\\
(N,n) \models r,  \hbox{\ if \ } r\in \cal Q.
\end{cases}
\]
Then the projection over the first   component is  a ${\cal P}$-bisimulation between $K$ and $M$, while the  projection over the second   component is   a ${\cal Q}$-bisimulation between $K$ and $N$.
\end{lemma}
 
We next show that the amalgamation property of Kripke models extends easily to team models:
\begin{lemma}\label{team_amalgamation} If ${\cal P}, {\cal Q}$ are sets of propositions and $(M,X)$, $(N,Y)$ are team models such that 
\begin{equation}\label{pcapq}
(M,X)\rightleftharpoons_{{\cal P}\cap {\cal Q}}(N,Y)
\end{equation}
 then there is a  team model $(K,Z)$ such that 
\[(M,X)\rightleftharpoons_{\cal P }(K,Z)\rightleftharpoons_{\cal Q }(N,Y)\] 
\end{lemma}
{\bf Proof}
Let $B$ be the bisimulation  witnessing  $(M,X)\rightleftharpoons_{ {\cal P}\cap {\cal Q} } (N,Y)$ as in Remark \ref{max}; consider  the $B$-amalgamation  $K$  of $M,N$ as defined   in \ref{kripke_amalgamation}.   

We define $Z=(X\times Y)\cap B$; if $x\in X$ then, since $B$ is a ${\cal P}\cap {\cal Q}$-bisimulation between the team models $(M,X)$ and $(N,Y)$,   there exists $y\in Y$ such that   $(x,y)\in B$. Hence,  the pair $(x,y)$ belongs to $Z$,  and  the projection over the first   component is   a  witness fort a ${\cal P}$-team bisimulation between $(K,Z)$ and $(M,X)$.
Similarly  the projection over the first   component is   a  witness fort a ${\cal Q}$-team bisimulation between $(K,Z)$ and   $(N,Y)$.
\finishproof

\section{Bisimulation Quantifiers and Uniform Interpolation in Modal Team Logic}

Given  a    logic $L$ with modal team semantics such that  all formulas  are bisimulation invariant,  we  extend  its  syntax by means of the {\em existential bisimulation   quantifier}, $\tilde \exists p~\phi$,  obtaining the logic 
$\tilde \exists L$. E.g.  a formula $\phi$ in $\tilde \exists \mathcal {ML} $ is defined by:
\[\phi:=  p ~|~\neg p ~|~ \bot ~|~\phi_1\wedge\phi_2~|~ \phi_1\otimes \phi_2~|~ 
\Diamond\phi ~|~ \Box\phi~|~ \tilde \exists p ~\phi. \]

The semantics of $\tilde \exists p ~ \phi$ is defined as follows:   for any team model $(M,X)$ over a set of proposition $\cal P$ and for any $p\in {\cal P}$ it holds:
\begin{equation}\label{semantics}
\noindent  (M,X)\models \tilde \exists  p~ \phi \Leftrightarrow 
   \end{equation}  
  \[~~~~~~~~~~~\exists (M',X' )  \rightleftharpoons_{Free(\phi)\setminus \{p\}}  (M,X)~ \hbox{and} ~(M',X') \models \phi,\]
where the set of free variables $Free(\phi)$ of a formula $\psi\in \tilde\exists L$ are defined inductively as expected,  stipulating that
 $ Free(\tilde\exists p ~\phi)=Free(\phi)\setminus\{p\}$. 
One can easily prove that all formulas in $\tilde \exists L$ are   bisimulation invariant:

\begin{lemma} Suppose  all formula of $L$ are bisimulation invariant, $\phi\in \tilde \exists L$, $(M,X)\models \phi$, and $(M,X)  \rightleftharpoons_{Free(\phi)}(N,Y)$ then $(N,Y)\models \phi$. 
\end{lemma}
Moreover,  existential bisimulation quantifiers   in   $\tilde \exists L$  are related to uniform interpolants as follows:

\begin{lemma}\label{bisversusint} Consider a modal team logic $L$, invariant under bisimulation,  and let $\tilde\exists L$ be  its existential bisimulation  extension.   If $\phi$ is  a formula of   $\tilde\exists L$   then $\tilde \exists p~ \phi$ is  a uniform interpolant  for the formula $\phi$ in $\tilde \exists L$, with respect to  $Free(\phi)\setminus \{p\}$.
\end{lemma}
{\bf Proof} 
It is clear that $\phi \models \tilde \exists p ~\phi$ and, by definition,  $Free(\tilde\exists p ~\phi)=Free(\phi)\setminus \{p\}$.\\ Suppose $\phi\models \psi$ with $\psi\in  \tilde \exists L$  and $Free(\phi)\cap Free(\psi)\subseteq Free(\phi)\setminus \{p\}$, that is, $p\not \in Free(\psi)$. 
We prove that $\tilde\exists p ~ \phi\models \psi$. If $(M,X)\models \tilde\exists p~ \phi$ then there exists $(M',X')$ such that 
\[(M',X')\rightleftharpoons_{Free(\phi)\setminus \{p\}}  (M,X)\] and  $(M',X')\models \phi$.   Since  $Free(\phi)\cap Free(\psi)\subseteq Free(\phi)\setminus \{p\}$, by the amalgamation property  proved in Lemma \ref{team_amalgamation},   there exists  a team model $(N,Y)$ such that
\[(M',X')\rightleftharpoons_{{Free(\phi)}}  (N,Y)\rightleftharpoons_{{Free(\psi)}}(M,X).  \]
Since $(M',X')\models \phi$, the first bisimulation  implies $ (N,Y)\models \phi$. Then, from $\phi\models\psi$ we obtain  $ (N,Y)\models \psi$, and from  $(N,Y)\rightleftharpoons_{{Free(\psi)}}(M,X)$ we finally have $(M,X)\models \psi$. 
 \finishproof
 
 Lemma \ref{bisversusint} allows to use, in the Modal Team context, the well known strategy that consists on  proving  uniform interpolation in a logic $L$ by showing  that, for any formula $\phi\in L$ there is a formula $\theta\in L$ which is equivalent to $\tilde \exists p~ \theta$. 
Our first  task is   to use this strategy to  prove uniform interpolation for Full Modal Team Logic   $\mathcal {FMTL}$.  Notice that we cannot apply  Theorem \ref{Kontinen} directly, proving that for all $\phi\in \mathcal{FMTL}$ 
the formula $\tilde\exists p~ \phi \in  \mathcal{FMTL}$, because, although we proved that the property expressed by $\tilde\exists p~ \phi $ is bisimulation invariant, we do not know whether it is an $FO$-property.

In the following  we prove that the existential  bisimulation quantifier commutes with both  disjunctions $\otimes$, $\vee$, and  with  the non-emptyness atom $NE$.
First we prove that the singleton  semantics of the bisimulation quantifier over classical modal formulas  is equivalent to its team semantics.

\begin{lemma} \label{exmodal}
Suppose  $\theta,\phi \in  \mathcal {ML}$, and $\theta$ behaves as $\tilde\exists p~ \phi$  w.r.t. singleton  semantics, that is,    for all Kripke models $(M,w)$ it holds 
\[(M,w)\models \theta \Leftrightarrow  \exists  (N,v)   \rightleftharpoons_{{\cal L}(\phi)\setminus \{p\}} (M,w)~ \hbox{and  }~(N,v)\models \phi;\]
then  $\theta$ is equivalent to $\tilde \exists p ~\phi$ in the modal team semantics, that is, for all team models  $(M,X)$ it holds 
\[(M,X)\models \theta \Leftrightarrow  \exists  (N,Y)   \rightleftharpoons_{{\cal L}(\phi)\setminus \{p\}} (M,X)~ \hbox{and  }~(N,Y)\models \phi;\]
\end{lemma}
 {\bf Proof} 
 If $(M,X)\models \theta$    then, since $\theta \in \mathcal {ML}$,   for all $w\in X$ we have $(M,w)\models \theta$;    hence, for all $w\in X$ there exists $(N_w,v_w)$ such that 
 \[ (N_w,v_w)   \rightleftharpoons_{{\cal L}(\phi)\setminus \{p\}} (M,w)~ \hbox{and  }~(N_w,v_w)\models \phi.\]
 Consider the disjoint union $N$ of all the $N_w$,  for $w\in X$, and the team $Y=\{v_w:w\in X\}$. 
Since  $\phi$ is a classical modal formula,  we have $(N,Y)\models \phi$ and $(N, Y) \rightleftharpoons_{{\cal L}(\phi) \setminus \{p\}} (M,X)$, so that $(M,X)\models \tilde \exists p~ \phi$.  Hence $\tilde \exists p \phi$ is a logical consequence of $\theta$ in the modal team semantics. 

Vice versa,  if  $(M,X)\models \tilde \exists p~ \phi$ we prove that $(M,w)\models \theta$, for all $w\in X$.
 Let $(N,Y)$ be such that $(N,Y)\models \phi$ and $(N, Y) \rightleftharpoons_{{\cal L}(\phi) \setminus \{p\}} (M,X)$. 
Since $\phi$ is a classical modal formula, $(N,y)\models \phi$, for all $y\in Y$. If $w\in X$ then there exists $v\in Y$ such that  $(N,v)   \rightleftharpoons_{{\cal L}(\phi)\setminus \{p\}} (M,w)$  and
hence, by hypothesis $(M,w)\models \theta$. 

Finally,  from $(M,w)\models \theta$, for all $w \in X$, it follows $(M,X)\models \theta$, since $\theta$ is a classical modal formula. 
\finishproof

\begin{corollary}\label{classicalbexists}
If $\phi \in  \mathcal {ML}$ then $\tilde \exists p~ \phi \in \mathcal {ML}$.
\end{corollary}
 {\bf Proof}  ~The corollary follows  from the previous Lemma and the closure of  Classical Modal Logic under the existential bisimulation quantifier w.r.t.  singleton  semantics: for any formula $\phi$ of classical modal logic, there exists a formula $\theta$ such that   for all Kripke models $(M,w)$ it holds 
\[(M,w)\models \theta \Leftrightarrow  \exists  (N,v)   \rightleftharpoons_{{\cal L}(\phi)\setminus \{p\}} (M,w)~ \hbox{and  }~(N,v)\models \phi\] (for a proof see   \cite{Visser1996}). 
\finishproof 

Finally, we prove a  lemma stating that in the team semantics of $\tilde \exists p ~\phi$ we can substitute $Free(\phi)$ by any  set of propositions  containing the free variable of $\phi$. 
\begin{lemma}\label{globalex} If $\phi$ is a formula of $\mathcal{MTL}$, $(M,X)$ is a Kripke model over a set $Prop$ of propositional variables containing the free variables of $\phi$, 
and 
$(M,X)\models \tilde \exists p~ \phi$,  then  there exists  a model $ (K,Z) $ such that:
\[(K,Z) \rightleftharpoons_{Prop\setminus \{p\}} (M,X)~ \hbox{and  }~(K,Z)\models \phi.\]
\end{lemma}
{\bf Proof} If $(M,X)\models \tilde \exists p~ \phi$ then by definition there exists  a model $ (N,Y) $ such that:
\[(N,Y) \rightleftharpoons_{Free(\phi)\setminus \{p\}} (M,X)~ \hbox{and  }~(N,Y)\models \phi\]
Since $Free(\phi)\setminus \{p\}= Free(\phi) \cap (Prop\setminus \{p\})$, by Lemma \ref{team_amalgamation}  there exists a team model $(K,Z)$ with 
\[(N,Y) \rightleftharpoons_{Free(\phi)} (K,Z)~~ \hbox{and}~~ (M,X)\rightleftharpoons_{Prop\setminus\{p\}} (K,Z)\]
Then, since $(N,Y)\models \phi$,  using  the first bisimulation we obtain $(K,Z)\models \phi$, and  the lemma follows. 
\finishproof

\begin{lemma}\label{comm}
If $\phi_1, \phi_2\in   \mathcal {MTL} $  then 
\[\tilde \exists p ~(\phi_1\wedge NE)\equiv( \tilde  \exists p ~\phi_1) \wedge NE\]
\[\tilde \exists p ~(\phi_1\lor \phi_2)\equiv \tilde \exists p~\phi_1 \lor \tilde \exists p~\phi_2\]
\[\tilde \exists p~(\phi_1\otimes \phi_2)\equiv\tilde \exists  p~\phi_1\otimes  \tilde \exists p~ \phi_2\]
\end{lemma} 
{\bf Proof}  The first equivalence holds because if the team of a team model in not empty, so is any team of a bisimilar team model. 

 We prove the third equivalence, leaving the second one  to the reader. 
 Let $(M,X)$ be a tem model.  
If $(M,X)\models \tilde \exists p~(\phi_1\otimes \phi_2)$  there exists  a team model $(N,Y)$ such that $(N,Y)\models  \phi_1\otimes\phi_2$ and 
\[(N,Y ) \rightleftharpoons_{Free(\phi_1\otimes \phi_2)\setminus \{p\}}  (M,X).\]  Then,  there are $Y_1,Y_2$ with $Y=Y_1\cup Y_2$ and 
$(N,Y_i)\models  \phi_i$, for $i=1,2$. 
Let $X_i=\{x\in X :\exists y\in Y_i ~ (N,y) \rightleftharpoons_{Free(\phi_1\otimes \phi_2)\setminus \{p\}} (M,x)\}$. 
Then $X=X_1\cup X_2$ and 
\[(N,Y_i) \rightleftharpoons_{Free(\phi_1\otimes \phi_2)\setminus \{p\}}   (M,X_i).\] Since $Free(\phi_1\otimes \phi_2) \supseteq Free(\phi_i )$ we  also have 
\[(N,Y_i) \rightleftharpoons_{Free(\phi_i)\setminus \{p\}}   (M,X_i), \]  and from $(N,Y_i)\models  \phi_i$ it follows $(M,X_i)\models ~\exists p ~\phi_i$,  for $i=1,2$.  This implies $(M,X)\models  \tilde \exists p~\phi_1\otimes  \tilde \exists p~ \phi_2$. 

Conversely, suppose  $(M,X)\models  \tilde \exists p ~\phi_1\otimes  \tilde \exists p~ \phi_2$. Then $X=X_1\cup X_2$ with $(M,X_i)\models  \tilde \exists p~ \phi_i$, for $i=1,2$. 
 Using  Lemma \ref{globalex} we obtain models $(K_i,Z_i)$ such that  $(K_i,Z_i)\models \phi_i$ and 
 \[(K_i,Z_i)  \rightleftharpoons_{Free(\phi_1\otimes \phi_2)\setminus \{p\}}   (M,X_i).\]
Define $(N,Z):= (K_1\dot{\cup} K_2, Z_1\dot{\cup} Z_2 )$, where $\dot{\cup}$ denotes disjoint union. For all $y\in Z_i$ we have 
  \[(N,y) \rightleftharpoons_{Free(\phi_1\otimes \phi_2)}  (K_i,y),\] hence 
 \[(N,Z_i) \rightleftharpoons_{Free(\phi_1\otimes \phi_2)}  (K_i,Z_i).\]  It follows that  $(N,Z_i)\models \phi_i$, hence $(N,Z)\models \phi_1\otimes \phi_2$  and 
\[(N,Z) \rightleftharpoons_{Free(\phi_1\otimes \phi_2)\setminus \{p\}}  (M,X).\]This implies $(M,X)\models \tilde \exists p ~(\phi_1\otimes \phi_2)$.   
\finishproof

Using the   previous lemmas  we can now prove that $\tilde\exists   \mathcal {MTL} $ is expressively equivalent to $ \mathcal {MTL}$, but first we need to fix some notation and recall some well known result of 
characteristic formulas for modal logic. 
 If $(M,w)$ is a Kripke model and $\cal P$ is a finite set of propositions we denote by 
  $ \phi^{k, \cal P}_{M,w}$ the modal formula characterizing $(M,w)$ modulo $k$-bisimulation w.r.t. the variables in $\cal P$, so  that,  for all Kripke models $(M',w')$ it holds:
  \[ (M',w')\models  \phi^{k, \cal P}_{M,w} ~~\Leftrightarrow~~ (M',w') \rightleftharpoons^k_{\cal P} (M,w).\]
  We omit the reference to $\cal P$ if this set is clear from the context 
  (for a definition of  $ \phi^k_{M,w} $ see e.g. \cite{Goranko}).
\begin{theorem}  \label{bisq} Elimination of bisimulation quantifiers in $ \mathcal {MTL}$: for any $\phi\in \mathcal {MTL}$ and $p\in \cal P$ there exists a formula  $\theta \in \mathcal {MTL}$ which is equivalent to $ \tilde \exists p~ \phi$. 
\end{theorem}
{\bf Proof} Given a team model  $(M,X)$, for all $w\in X$  consider the characteristic formulas $\chi^k_{(M,w)}$ with respect to $\cal L(\phi)$ and    define \[\chi^k_{(M,X)}:=  \bigotimes_{w\in X} (\chi^k_{(M,w)}\wedge NE).\]   
Then,   as proved in \cite{Kontinen2015}, for all models $(N,Y)$ it holds:
\[(N,Y)\models  \chi^k_{(M,X)} ~\Leftrightarrow~ (N,Y) \rightleftharpoons^k_{\cal L(\phi)}  (M,X).\]   This implies that any formula 
  $\phi\in \mathcal {MTL}$  of modal  depth  $k$ is equivalent to a disjunction of formulas  $\chi^k_{(M,X)}$:
  \[\phi \equiv  \bigvee_{(M,X)\models \phi} \chi^k_{(M,X)}.\]
By Lemma  \ref{comm} we have 
  \[\tilde \exists p ~\phi \equiv \bigvee_{(M,X)\models \phi}\tilde  \exists p ~ \chi^k_{(M,X)} \equiv  \bigotimes_{w\in X} ( \tilde \exists p ~\chi^k_{(M,w)} \wedge NE).\]

By Corollary \ref{classicalbexists}, the formulas $ \tilde \exists p ~\chi^k_{(M,w)}$ are equivalent to  modal formulas and the theorem follows. 
  
\begin{corollary}  The team logic $\mathcal {MTL}$ enjoys uniform interpolation. 
\end{corollary}
{\bf Proof}
This follows from the previous theorem and Lemma \ref{bisversusint}.
\finishproof
\subsection{Uniform Interpolation for Modal Team Fragments.}
 
 In this section we prove uniform interpolation for all   fragments of $\cal MTL$ described in  Section  \ref{modal_fragments}, with one notable exception: Modal Independence Logic and its propositional fragment. 
 
 If $\mathcal C$ is a  team model  property    and $Q\subseteq Prop$,   we denote by  $\mathcal C_{\sim Q}$  the team  model property 
 of all team models which are in $\mathcal C$  modulo bisimulations forgetting  the variables in $Q$:
 
  \[ \mathcal C_{\sim Q}=\{ (N,Y): \exists (M,X) \in  \mathcal C ~~ (N,Y) \rightleftharpoons_{ Prop  \setminus Q}(M,X)\}.\]
  
  A collection of   team model properties is said to be {\em forgetting} if it is closed under the previous construction with respect to finite sets of propositions:

 \begin{definition}\label{forgetful_modal_class}
 A collection  $\cal T$   of  team model properties    is said to be {\em forgetting} if for every finite  $Q\subseteq Prop$
 and  $\mathcal C\in  \cal T$ it holds:
 \[\mathcal C_{\sim Q}\in \cal T.\] \\
 A fragment $L$ of  $\mathcal{MTL}$ is said to be forgetting if  $Team_L$ is so.\end{definition} 
  
  If $L$ is a team modal logic, $\phi\in L$, and  $Q=\{p_1, \ldots, p_n\}$   is a finite set of propositional variables, then    using Lemma \ref{globalex}  we get:
 \[||\tilde \exists p_1\ldots \tilde \exists p_n ~ \phi||=||\phi||_{\sim  Q}\] because 
  \[||\tilde \exists p_1\ldots \tilde \exists p_n ~ \phi||=\]
 \[=\{(N,Y) : \exists (M,X) \in  ||\phi||  \hbox{~with~}   (N,Y) \rightleftharpoons_{ \mathcal {L}(\phi)  \setminus Q} (M,X)\}=\]
 \[\{(N,Y) : \exists (M,X) \in  ||\phi||  \hbox{~with~}   (N,Y) \rightleftharpoons_{ Prop \setminus Q} (M,X)\} =||\phi||_{\sim  Q}. \] 
 
This easily implies:

 \begin{lemma} \label{forgetting} If  $L$  is a fragment   of $\mathcal {MTL}$ then 
 \begin{center}
$L$ is forgetting ~~$\Leftrightarrow$~~  $L$  enjoys uniform interpolation. 
 \end{center}
 \end{lemma}
 
 {\bf Proof} $(\Rightarrow) $~~ If $\phi\in L$ and $p_1, \ldots, p_n$  are    propositional variables  consider the set $Q=  \{p_1, \ldots, p_n\}$.  By the forgetting hypothesis, since $||\phi||\in Team_L$ then   
 $||\phi||_{\sim Q}= ||\tilde \exists p_1\ldots \tilde \exists p_n ~ \phi|| \in Team_L$. 
Hence,    $\tilde \exists p_1\ldots \tilde \exists p_n~ \phi$    is equivalent to a formula $\theta$ in $L$.     By   Lemma  \ref{bisversusint} we know that $\theta$  is a uniform interpolant for $\phi$ in $\tilde \exists L$ and, even more so, in $L$. 

$(\Leftarrow) $~~ 
If  $\phi\in L$  and $Q=\{p_{1}, \ldots, p_{n}\}$, then   the class of  team models $||\phi||_{\sim Q}$ is expressible by the  formula $\tilde \exists p_{1}\ldots \tilde \exists p_{n} ~\phi $ which,  by  Lemma \ref{bisversusint},  is a $\tilde \exists L$ uniform interpolant in  of the  $L$-formula $\phi$ w.r.t. ${\cal L}(\phi)\setminus Q$. Since $L$ enjoys uniform interpolation, 
and uniform interpolants are unique modulo equivalence, we obtain that  $\tilde \exists p_{1}\ldots \tilde \exists p_{n}~ \phi $ is equivalent to an $L$-formula and hence  $||\phi||_{\sim Q}\in Team_L$. 

 \finishproof

 \begin{lemma}
\label{closure} The following collections  of team model  properties are forgetting:
\begin{enumerate}
\item  the class of bisimulation invariant,  non empty,  downward closed and  union  closed (also called: flat) properties;
\item  the class of bisimulation invariant,  non empty,  downward closed  properties;
 \item  the class of bisimulation invariant,  non empty,  union  closed  properties. 
\end{enumerate}
\end{lemma}
{\bf Proof} ~ We prove the second and the third property, from which the first property  follows. 
\begin{enumerate}
\item Suppose $\mathcal C$  is  a non empty, bisimulation invariant,  downward closed set of teams, and   $Q$ is a finite set of propositions. We want to prove that 
\[\mathcal C_{\sim Q}=\{ (N,Y): \exists (M,X )\in  \mathcal C ~~  (N,Y) \rightleftharpoons_{ Prop \setminus Q} (M,X) \}  \] is non empty,  bisimulation invariant and downward closed.
The first two properties are easily verified and we leave them to the reader. As for the third property, suppose $(N,Y)\in \mathcal C_{\sim Q}$, and $Y'\subseteq Y$.  
 Then there exists $(M,X)$ with 
\[(N,Y) \rightleftharpoons_{Prop \setminus Q} (M,X)  ~~ \hbox{and}~~ (M,X)  \in  \mathcal C. \]
  For $s\in Y'$, let $s'\in X $ be such that  $(N,s') \rightleftharpoons_{Prop  \setminus Q} (M,s)$;   let $X'\subseteq X$ be
$X'=\{s'~:~s\in Y'\}$.  
By downward closure of $\mathcal C$, we have $(M,X')\in \mathcal C$; since  
\[(N,Y') \rightleftharpoons_{Prop  \setminus Q} (M,X')\]
we obtain $(N,Y')\in  \mathcal C_{Q}$. 

\item If $\mathcal C$  is  a  non empty, bisimulation invariant, union  closed set of teams over  $Prop$ and $Q$ is a finite set  $Q\subseteq Prop$, we want to prove that 
\[\mathcal C_{\sim Q}=\{ (N,Y): \exists (M,X )\in  \mathcal C ~~  (N,Y) \rightleftharpoons_{ Prop \setminus Q} (M,X) \}  \] is non empty,  bisimulation invariant,  and  union  closed.
The first two properties are easily verified and we leave them to the reader. As for the third property, 
suppose $(N,Y_1), (N,Y_2)\in \mathcal C_{\sim Q}$. Then  there are teams $(M_1,X_1), (M_2,X_2)\in \mathcal C$ such that, for $i=1,2$ it holds:
\[(N,Y_i) \rightleftharpoons_{Prop \setminus Q} (M_i,X_i)  ~~ \hbox{and}~~ (M_i,X_i)  \in  \mathcal C.\]
 Let $M$ be the disjoint union of $M_1, M_2$, and let $X$ be the disjoint union of $X_1, X_2$.  Then one can easily prove that, for $i=1,2$ it holds:
 \[(M,X_i) \rightleftharpoons_{Prop } (M_i,X_i), ~~\hbox{and}~~ (M,X_i\cup X_2 ) \rightleftharpoons_{Prop  \setminus Q} (N,Y_1\cup Y_2).\]
From the first two bisimulations we obtain $(M,X_i)\in \mathcal C$, which implies $(M,X_i\cup X_2 )$ since $\mathcal C$ is closed under union. 
It follows that $(N,Y_1\cup Y_2)\in \mathcal C_{\sim Q}$.
\end{enumerate}
\finishproof

Using Lemma \ref{closure}  and Lemma \ref{forgetting} we easily obtain:

\begin{corollary} The following fragment of  $\mathcal{MTL}$ enjoys uniform interpolation:
\[ \mathcal{ML},~~ \mathcal{MDEP}, ~~\mathcal{MINC}\]
\end{corollary}

Finally, we consider propositional fragments.  Being complete for the class of all team properties, it is clear that $\mathcal{FPT}$ enjoys uniform interpolation. Then we can use the 
semantical characterization of the fragments $\mathcal{CPL},~~\mathcal{PD},~~\mathcal{PINC}$ to prove, as we did for modal fragments, that all these logics enjoy uniform interpolation. 
 
 \begin{corollary} The following fragments of  $\mathcal{FPT}$ enjoys uniform interpolation:
\[\mathcal{CPL},~~\mathcal{PD},~~\mathcal{PINC}\]
\end{corollary}
\subsection{Conclusion and Open Questions}

The method above allow us  to prove uniform interpolation for  propositional and modal  team logics  whose  class of team models  is {\em forgetting} (see Def. \ref{forgetful_modal_class}). However, to prove that the class of team models for a logic is forgetting,  we used  a good description  of the class,  given in  Theorem \ref{Kontinen}. To our knowledge, this description is missing for  indipendence logic, for which uniform interpolation remains an open question.

\end{document}